\documentclass[aps,prd,twocolumn,showkeys,floats,showpacs, nofootinbib]{revtex4-1}
\usepackage{graphics,epsfig,color} 
\usepackage{varioref,exscale,latexsym,amsmath,amssymb}
\usepackage{hyperref}
\usepackage{slashed}

\def\L {\mathcal{L}}

\def\be{\begin{equation}}
\def\ee{\end{equation}}
\def\bea{\begin{eqnarray}}
\def\eea{\end{eqnarray}}
\begin{document}
\title{Merging matter and geometry in the same Lagrangian}
\author{Hendrik Ludwig}
\affiliation{ICRANet, Piazza della Repubblica 10, I-65122 Pescara, Italy}
\affiliation{Dipartimento di Fisica and ICRA, Sapienza Universita
di  Roma Placia Aldo Moro 5, I-00185 Rome, Italy}
\affiliation{UMR ARTEMIS, CNRS, University of Nice Sophia-Antipolis,
Observatoire de la C\^ote d’Azur, BP4229, 06304, Nice Cedex 4, France}
\author{Olivier Minazzoli}
\email[Corresponding author: ]{ominazzoli@gmail.com}
\affiliation{Centre Scientifique de Monaco, 8 Quai Antoine 1er, Monaco}
\affiliation{UMR ARTEMIS, CNRS, University of Nice Sophia-Antipolis,
Observatoire de la C\^ote d’Azur, BP4229, 06304, Nice Cedex 4, France}
\affiliation{ICRANet,  University  of  Nice-Sophia  Antipolis,
28  Av.   de  Valrose,  06103  Nice  Cedex  2,  France}
\author{Salvatore Capozziello}
\affiliation{Dipartimento di Fisica, Universit\`a di Napoli ``Federico II'', Compl. Univ. di Monte S. Angelo, Edificio G,Via Cinthia, I-80126, Napoli, Italy}
\affiliation{Istituto Nazionale di Fisica Nucleare (INFN),  Sezione di Napoli, Compl. Univ. di Monte S. Angelo, Edificio G, Via Cinthia, I-80126, Napoli, Italy}
\affiliation{Gran Sasso Science Institute (INFN), Viale F. Crispi, 7, I-67100 L'Aquila, Italy. }
\begin{abstract}
We  show that a Lagrangian density proportional to $\sqrt{-g} \L_m^2/R$ reduces to a pressuron theory of gravity that is indistinguishable from General Relativity in the dust limit. The combination of matter and geometry in the same Lagrangian density intrinsically satisfies Mach's Principle --- since matter cannot exist without curvature and vice versa --- while it may have the correct phenomenology in order to describe actual gravity.
\end{abstract}

\maketitle

{\it Introduction}.---  Several issues point out toward potential inconsistencies of General Relativity at ultraviolet and infrared scales. Specifically, while this theory perfectly works at solar system scale, shortcomings appear at quantum, astrophysical and cosmological levels. Quantum Gravity is the main conundrum as General Relativity cannot be dealt within standard Quantum Field Theory (QFT). On the other hand, we need dark energy and dark matter to achieve a self-consistent picture for the future standard model of cosmology. Up to now, no new particle beyond the current Standard Model of particles has been detected, while a huge amount of missing matter and a cosmic speed up are needed to fit the observed dynamics.
 
These issues can be softened by taking into account alternative theories of gravity, in particular scalar-tensor and higher-order gravity \cite{fujii:2003fi,*woodard:2007ln,*defelice:2010lr,*sotiriou:2010rp,*capozziello:2011pr,*clifton:2012fk,*capozziello:2015sc} --- that are semi-classical models where additional degrees of freedom are introduced, hence enlarging the dynamics of General Relativity. The method consists in adding minimally and/or non-minimally scalar fields and/or higher-order curvature invariant and see whether or not it helps to soften the issues of General Relativity. The presence of these terms can also be justified by perturbative QFT since matter-gravity interactions on perturbatively curved spacetimes result in such corrections \cite{birrell:1984bk}. 
 
From the infrared perspective, one of the goals is to address the dark side with these additional gravitational degrees of freedom \cite{capozziello:2008gr, capozziello:2012ap,nojiri:2004pl,*bertolami:2007pd,*harko:2008pl,*bertolami:2008uq,*sotiriou:2008cq,*defelice:2010lr,*nojiri:2011pr,*bertolami:2012fl,*wang:2012cq,*tamanini:2013pr,*harko:2014ga}. Indeed these theories can be re-expressed effectively as the theory of Einstein plus additional effective terms entering the right hand side of the metric field equation. Therefore, in some sense, they can be considered as effective source terms like standard matter. For instance, a commonly considered straightforward extension of General Relativity is $f(R)$ gravity where the curvature term in the action is an algebraic function of the undifferentiated Ricci scalar, while the matter part is left unchanged.

However, because of the current tight constraints one has on gravity \cite{will:2014lr}, any proposed alternative theory of gravity must be such that it behaves like General Relativity in most situations \cite{damour:1993uq,*damour:1993kx,*khoury:2004uq,*khoury:2004fk,*khoury:2010ax,*defelice:2010lr,*sotiriou:2010rp,*capozziello:2011pr,*brax:2013cq}. Driven by this observation, and having in mind solar system experiments and constraints in particular, Minazzoli and Hees \cite{minazzoli:2013fk,*minazzoli:2014ao,*minazzoli:2014pb,minazzoli:2015ax} recently proposed another way to allow some scalar-tensor theories to satisfy experimental and observational constraints. Indeed, with a specific scalar-matter coupling scalar-tensor theories reduce to General Relativity in weak pressure regimes. For that reason the specific scalar-field(s) associated with this kind of theory has been dubbed ``\textit{pressuron(s)}'' \cite{minazzoli:2013fk,*minazzoli:2014ao,*minazzoli:2014pb,minazzoli:2015ax}.

On the other side, there is a well-known equivalence between $f(R)$ theories and a sub-class of scalar-tensor theories when the matter Lagrangian is minimally coupled to the gravitational field(s). Hence, one can expect a similar equivalence between a pressuron theory and a $f(R)$ theory with non-minimal gravity-matter coupling. Such non-minimal couplings have recently been investigated, notably in \cite{nojiri:2004pl,*bertolami:2007pd,*harko:2008pl,*bertolami:2008uq,*sotiriou:2008cq,*defelice:2010lr,*nojiri:2011pr,*bertolami:2012fl,*wang:2012cq,*tamanini:2013pr,*harko:2014ga}. They have been extended to $f(R,\L_m)$ and $f(R,\L_m,\phi,(\partial \phi)^2)$ theories later on \cite{harko:2010ep,*harko:2013rz}.

In what follows we present a simple and elegant $f(R,\L_m)$ theory that turns to be equivalent to a special case of pressuron theories. As we will see, the main surprise comes from the fact that the whole Lagragian is described by only one term where matter and geometry are related by a multiplicative coupling.\\

{\it On an unexpected action form}.---  The action of this theory can be set as  follows
\be
S=-\frac{1}{2} \int d^4x \sqrt{-g} \kappa \frac{\L_m^2}{R}, \label{eq:actionfR}
\ee
with $\kappa \equiv 8\pi G/c^4$, $G$ and $c$ are the gravitational constant \footnote{$G$ is different from the effective gravitational constant that is measured in Cavendish experiments. The effective gravitational constant can be deduced in this theory by doing a post-Newtonian expansion of the field equations --- see for instance \cite{minazzoli:2013fk,*minazzoli:2014ao,*minazzoli:2014pb}.} and the speed of light respectively. One has to stress that action (\ref{eq:actionfR}) is the complete action of the theory. 

First of all, note that in this theory, the dynamics can only exist if there is matter (ie. $\L_m \neq 0$). This would suggest that there is a deeper link between space-time and matter than usually assumed. In particular, Mach's principle is intrinsically fulfilled in this theory. This fact will be even more obvious later on after we rewrite the theory in its scalar-tensor form \footnote{Let us also remind that scalar-tensor theories were originally motivated by Mach's principle \cite{brans:1961fk,*brans:2014sc}.}. But consequently, it also means that the theory does not exist in vacuum. Fortunately, QFT tells us that true vacuum does not exist because of 0-point energy. Hence, action (\ref{eq:actionfR}) may actually be well defined everywhere, representing a realistic interplay between matter and geometry. Another thing to point out is that action (\ref{eq:actionfR}) is one of the simplest choices with the correct dimension that involves a multiplicative coupling between curvature and matter rather than an additional coupling. 

The metric field equation of the theory writes
\be
R_{\mu \nu}-\frac{1}{2} g_{\mu \nu} R= - \frac{R}{\L_m} T_{\mu \nu} + \frac{R^2}{\L_m^2} \left(\nabla_\mu \nabla_\nu - g_{\mu \nu} \Box \right)\frac{\L_m^2}{R^2}, \label{eq:fRmetricfield}
\ee
with the stress-energy tensor defined as follows
\be
T_{\mu \nu} \equiv - \frac{2}{\sqrt{-g}} \frac{\delta \left(\sqrt{-g} \L_m \right)}{\delta g^{\mu \nu}}. \label{eq:defSET}
\ee
The trace of the metric field equation reads
\be
3 \frac{R^2}{\L_m^2} \Box \frac{\L_m^2}{R^2}=  R-\frac{R}{\L_m} T. \label{eq:tracemetricFE}
\ee
Let us remind that for pressureless perfect fluids, one has $\L_m=-c^2 \rho=T$, \footnote{$\rho=\sum_A (\sqrt{-g} U_A^0)^{-1} m_A \delta^{(3)}(x^\alpha-x_A^\alpha)$, where $ U_A^\alpha \equiv dx^\alpha/c d \tau_A$, $m_A$ is the conserved mass of a particle such that $dm_A/d\tau_A=0$ and $\delta^{(3})$ the Dirac distribution in three dimensions.} where $c^2 \rho$ is the rest mass energy density \cite{harko:2010pr,*minazzoli:2012pr,*minazzoli:2013zl}. Therefore, the right hand side of equation (\ref{eq:tracemetricFE}) is null for pressureless perfect fluids. This is precisely the pressuron mechanism that has been described in \cite{minazzoli:2013fk,*minazzoli:2014ao,*minazzoli:2014pb,minazzoli:2015ax}.

It turns out that as in usual $f(R)$ gravity one can re-write the action in an equivalent scalar-tensor theory form. Indeed, defining
\be
\sqrt{\Phi}=h=- \kappa \frac{\L_m}{R},\label{eq:defscalar}
\ee
the field equations can be re-written as follows
\bea
&&R_{\mu \nu}-\frac{1}{2} g_{\mu \nu} R= \frac{\kappa}{h} T_{\mu \nu}+ \frac{1}{h^2}\left(\nabla_\mu \nabla_\nu - g_{\mu \nu} \Box \right) h^2, \label{eq:fEscalarmertric} \\
&&\frac{3}{h^2} \Box h^2= \frac{\kappa}{h} \left(T-\L_m \right).\label{eq:fEscalarscalar}
\eea
But such field equations can be derived from the following effective scalar-tensor theory action
\bea
S&=&\int  d^4x \sqrt{-g} \left[\frac{\Phi R}{2 \kappa}+ \sqrt{\Phi} \L_m \right] \nonumber\\
&=&\int  d^4x \sqrt{-g} \left[\frac{h^2 R}{2 \kappa}+ h \L_m \right]. \label{eq:actionscalar}
\eea
Such an action is the one of a pressuron without kinetic terms (ie. $\omega(\Phi)=0$) \cite{minazzoli:2013fk,*minazzoli:2014ao,*minazzoli:2014pb,minazzoli:2015ax} \footnote{One can see that starting from action (\ref{eq:actionscalar}), one recovers the field equations (\ref{eq:fRmetricfield}-\ref{eq:tracemetricFE}) by inserting (\ref{eq:fEscalarscalar}) into (\ref{eq:fEscalarmertric}) and then re-expressing the scalar field equation without $h$. Hence actions (\ref{eq:actionfR}) and (\ref{eq:actionscalar}) are indeed totally equivalent as long as true vacuum energy does exist.}. But let us recall that even for $\omega=0$, pressuron theories satisfy the strong constraints one has from solar system experiments \cite{minazzoli:2013fk,*minazzoli:2014ao,*minazzoli:2014pb,minazzoli:2015ax}. For instance the post-Newtonian parameter $\gamma$ is exactly equal to one, just as in General Relativity \cite{minazzoli:2013fv}. In general, regardless the value of $\omega$, pressuron theories reduce to the General Relativity phenomenology in weak pressure regimes such as in the solar system or during the late cosmic period \cite{minazzoli:2013fk,*minazzoli:2014ao,*minazzoli:2014pb,minazzoli:2015ax}. Indeed, for a barotropic perfect fluid, the on-shell Lagrangian is minus the total energy density ($\L_m = - \epsilon$) \cite{harko:2010pr,*minazzoli:2012pr,*minazzoli:2013zl} and therefore equation (\ref{eq:fEscalarscalar}) reduces to \cite{minazzoli:2013fk,*minazzoli:2014ao,*minazzoli:2014pb,minazzoli:2015ax}
\be
\frac{1}{h^2} \Box h^2= \kappa \frac{ P}{h},
\ee
where $P$ is the barotropic pressure of the fluid \cite{harko:2010pr,*minazzoli:2012pr,*minazzoli:2013zl}, such that the scalar-field source disappears for $P=0$ --- in accordance with the dust case previously discussed. Another way to see the equivalence of actions (\ref{eq:actionfR}) and (\ref{eq:actionscalar}) goes as follows: using equation (\ref{eq:defscalar}), one can write
\be
-\frac{1}{2} \kappa \frac{\L_m^2}{R}= -\frac{a+b}{2} \kappa \frac{\L_m^2}{R}=\frac{a}{2} h \L_m-\frac{b}{2} h^2 \frac{R}{\kappa},
\ee
with $a$ and $b$ two constants such that $a+b=1$. Then from the definition of the stress-energy tensor (\ref{eq:defSET}) and in order to get the appropriate normalization of the material Lagrangian, one necessarily has $a=2$ \footnote{$a=2$ in order to recover for instance the fact that in the dust case limit one has $S_m=-c\int m h d\tau$. One can note that obviously, total inertia of massive particles depends on $h$ in this model, which depends on the content of the universe, therefore satisfying Mach's principle.}, therefore implying $b=-1$. Hence, one gets

\bea
-\frac{1}{2} \kappa \frac{\L_m^2}{R}= \frac{1}{2} h^2 \frac{R}{\kappa}+ h \L_m. \label{eq:equivalence}
\eea
At the same time this choice for $a$ and $b$ is unique in the sense that the variation of $h$ with respect to the metric cancels for the two terms in this action.
This actually allows to treat $h$ as a fundamental scalar field, while it is a priori dependent on the metric, and establishes the equivalence to the pressuron
with $\omega=0$.
\\
However there is a strong difference with usual pressuron theories. Indeed, in this theory there cannot be any exact vacuum solutions as the theory is not even defined in a vacuum configuration \footnote{According to \cite{pais:1982bk} (p.287), in 1917 Einstein himself believed ``that the correct equations [for gravity] should have no solutions at all in the absence of matter''. This is because Einstein believed in Mach's principle --- that states roughly speaking that ``total inertia of a mass point is an effect due to the presence of all other masses, due to a sort of interaction with the latter'' --- while vacuum solutions would mean that a test particle's inertia would be defined even without any other form of matter, therefore in contradiction with Mach's principle \cite{pais:1982bk}.} (see action (\ref{eq:actionfR})). However, as in General Relativity with a cosmological constant, vacuum solutions may be good approximations in some situations. But in general, one has to consider the contribution of the vacuum energy in the field equations from the very definition of the theory, even though the issue of its nonphysical \footnote{\textit{Nonphysical} in the sense that it seems to be different from what is observed.} value derived from the usual QFT pertubative technics has not been settled yet \cite{bianchi:2010na}. Hence, part of the stress-energy tensor (\ref{eq:defSET}) must come from vacuum energy, otherwise the theory would not even be defined. \\

{\it Conclusion}.--- In this communication, we presented a $f(R,\L_m)$ theory of gravity with a very unusual action that reduces to General Relativity in pressure-less regimes. Because matter and curvature no-longer couple additively in the action but multiplicatively (\ref{eq:actionfR}), curvature and matter are even more intrinsically related than in General Relativity, such that it seems that dynamics do not exist without matter. This fulfills Einstein's initial proposal of having a theory of gravity that satisfies Mach's principle \cite{pais:1982bk}. 

But let us stress that it comes from a change of paradigm. Indeed, curvature and matter are usually considered separately before their mutual effects are simply added up in the action. But here, curvature and matter are related multiplicatively in the action from the start. Therefore, in the present theory matter cannot be considered without curvature and vice versa. Hence, in some sense, one can see this theory as a unified theory of matter and geometry. As a consequence, the Planck mass and all particle masses are proportional in this theory (see equation (\ref{eq:equivalence})) \cite{minazzoli:2015ax}. However, it also means that the the 0-point energy value issue is even more pressing in this theory as the theory seems to be not even defined without considering such vacuum energy. 

Otherwise, we showed that this theory reduces to a special case of the so-called scalar-tensor pressuron theory. This relation may help in order to study it in various regimes (eg. cosmological, black hole physics etc.). In particular, thanks to the equivalence between the actions (\ref{eq:actionfR}) and (\ref{eq:actionscalar}), one can use results from the literature related to theories with multiplicative scalar-matter coupling (see for instance \cite{shifman:1978pb,*taylor:1988pb,*ellis:1989pb,*damour:1994uq,*damour:1994fk,*gasperini:2002kx,*damour:2002ys,*uzan:2011vn,*hees:2014pr,*hees:2015gg} and references therein). In any case, a lot of work is still needed in order to figure out whether or not the particular action (\ref{eq:actionfR}) is suitable to describe actual space-time dynamics --- down to the quantum level.

\end{document}